\documentclass[journal=nalefd,manuscript=letter]{achemso}
\usepackage[version=3]{mhchem} % Formula subscripts using \ce{}
\usepackage[T1]{fontenc}
\usepackage{textcomp}
\newcommand{\CPX}{\mathrm{CsPbX}_3}
\newcommand{\CPB}{\mathrm{CsPbBr}_3}
\newcommand{\IS}{I_\mathrm{S}}
\newcommand{\IPL}{I_{\mathrm{PL}}}

\newcommand{\QY}{\mathrm{PLQY}}
\newcommand{\Wcm}{\mathrm{W}/\mathrm{cm}^2}

\newcommand{\Pon}{P_\mathrm{on}}
\newcommand{\Poff}{P_\mathrm{off}}
\newcommand{\Ponoff}{P_\text{on(off)}}
\newcommand{\Sonoff}{S_\text{on(off)}}

\newcommand{\microsec}{\text{\textmu s}}
\newcommand{\ns}{\text{ns}}
\newcommand{\ps}{\text{ps}}

\renewcommand{\d}{\mathrm{d}}
\newcommand{\uPL}{\textmu-PL}

\author{Vigneshwaran Chandrasekaran}
\alsoaffiliation[Center for Nano and Biophotonics, Ghent University]{Center for Nano and Biophotonics, Ghent University, Belgium}
\author{Micka\"{e}l D. Tessier}
\alsoaffiliation[Center for Nano and Biophotonics, Ghent University]{Center for Nano and Biophotonics, Ghent University, Belgium}
\author{Dorian Dupont}
\alsoaffiliation[Center for Nano and Biophotonics, Ghent University]{Center for Nano and Biophotonics, Ghent University, Belgium}
\author{Pieter Geiregat}%
\alsoaffiliation[Center for Nano and Biophotonics, Ghent University]{Center for Nano and Biophotonics, Ghent University, Belgium}
\author{Zeger Hens}
\alsoaffiliation[Center for Nano and Biophotonics, Ghent University]{Center for Nano and Biophotonics, Ghent University, Belgium}
\author{Edouard Brainis}
\email{Edouard.Brainis@UGent.be}
\alsoaffiliation[Center for Nano and Biophotonics, Ghent University]{Center for Nano and Biophotonics, Ghent University, Belgium}

\affiliation[Physics and Chemistry of Nanostructures, Ghent University]
{Physics and Chemistry of Nanostructures, Ghent University, Ghent, Belgium}

\title[Single-Photon Emission by Colloidal InP/ZnSe Quantum Dots]
  {Nearly Blinking-Free, High-Purity Single-Photon Emission by Colloidal InP/ZnSe Quantum Dots}


\begin{document}

%\begin{tocentry}
%\begin{center}
%	\includegraphics{TOC.pdf}
%\end{center}
%\end{tocentry}

\begin{abstract}

Colloidal core/shell InP/ZnSe quantum dots (QDs), recently produced using an improved synthesis method, have a great potential in life-science applications as well as in integrated quantum photonics and quantum information processing as single-photon emitters. Single-particle spectroscopy of 10-nm QDs with 3.2-nm cores reveals strong photon antibunching attributed to fast (70-ps) Auger recombination of multiple excitons. The QDs exhibit very good photostability under strong optical excitation. We demonstrate that the antibunching is preserved when the QDs are excited above the saturation intensity of the fundamental-exciton transition. This result paves the way towards their usage as high-purity on-demand single-photon emitters at room temperature. Unconventionally, despite the strong Auger blockade mechanism,  InP/ZnSe QDs also display very little luminescence intermittency (``blinking''), with a simple \textit{on}/\textit{off} blinking pattern. The analysis of single-particle luminescence statistics places these InP/ZnSe QDs in the class of nearly blinking-free QDs, with emission stability comparable to state-of-the-art thick-shell and alloyed-interface CdSe/CdS, but with improved single-photon purity. 
\end{abstract}

Solid-state emitters providing single photons on-demand are an essential building block in photonic quantum technology,\cite{Aharonovich2016} which involves applications such as quantum cryptography,\cite{Beveratos2002,Heindel2012} quantum simulation,\cite{Broome2010,Tillmann2013} and quantum metrology.\cite{Motes2015} Currently, a diversity of material platforms are considered as single-photon emitters, which include  color centers or defects in crystalline hosts,\cite{Aharonovich2014,Morfa2012,Utikal2014} carbon nanotubes,\cite{Ma2015} transition metal dichalcogenides,\cite{He2015} and epitaxial\cite{Buckley2012,Patel2010,Sapienza2015} and colloidal\cite{Brokmann2004,Brokmann2004a,Pisanello2010,Pisanello2013} semiconductor nanocrystals or quantum dots (QDs). Among these, colloidal QDs stand out since they offer an extensive design freedom (both at the level of nanocrystal synthesis and device integration) as well as room temperature operation.\cite{Buckley2012} The emission wavelength of colloidal QDs can be readily tuned by changing their diameter, and they can be formed from several semiconductors. Moreover, colloidal synthesis methods offer an extensive control over the QD size and shape, and enable complex heterostructures to be formed. In addition, the resulting QD dispersion or inks can be readily combined with a multitude of technology platforms for further processing.

Room temperature photon antibunching in the luminescence of single colloidal QDs was first reported for CdSe/ZnS core/shell QDs and attributed to the highly efficient, non-radiative Auger recombination of multi-excitons.\cite{Michler2000,Lounis2000} This enabled the emission of a single photon to be triggered by a high-intensity excitation pulse, with a near-unity probability of photo-excitation and a high-purity single photon emission.\cite{Brokmann2004,Brokmann2004a} On the other hand, the further development of CdSe/ZnS QDs as single photon emitters stalled because of their intermittent fluorescence or blinking, which leads to an unpredictable succession of bright and dark periods.\cite{Nirmal1996,Lounis2000,Brokmann2004} In the case of CdSe/CdS core/shell QDs, it was shown that blinking can be suppressed by implementing sufficiently thick CdS shells.\cite{Chen2008,Mahler2008} However, this approach leads to a significant reduction of the Auger recombination rate, possibly linked to the formation of alloyed interfaces. The concomitant promotion of the radiative recombination of multi-excitons reduces photon anti-bunching, an effect that is the more pronounced the higher the biexciton photoluminescence quantum yield.\cite{Park2014} The same issue is faced by attempts to suppress blinking by speeding up radiative recombination through, for example, plasmonic effects.\cite{Yuan2009} Since accelerating radiative recombination shifts the balance between radiative and non-radiative Auger recombination of multi-excitons, blinking suppression by enhancing radiative recombination also results in a loss of purity of single photon emission.\cite{Naiki2011}

Currently, the focus of QD synthesis research is shifting to materials made from semiconductors that offer even better optical properties or have a composition non restricted by regulations, where caesium lead halide perovskites ($\CPX$, X=Cl,Br,I) and indium phosphide are the most notable examples. Rapidly after the first reports on their synthesis,\cite{Protesescu2015} single $\CPB$ nanocrystals were shown to exhibit photon anti-bunching and, in one study, reduced blinking.\cite{Park2015,Raino2016} However, as these nanocrystals showed significant photodegradation, it remains unclear whether $\CPB$ nanocrytals can provide single photons on demand. Opposite from $\CPB$ nanocrystals, InP QDs received little attention, if any, as possible single-photon emitters even if first syntheses were developed shortly after the first hot injection synthesis of CdSe QDs.\cite{micic1994} More recently, a new approach to form InP-based QDs has been introduced, based on aminophosphines as the phosphorous precursor.\cite{Song2013,Tessier2015} This method enables a variety of InP-based core/shell QDs to be formed with widely tunable optical properties. In this respect, the InP/ZnSe core/shell combination is interesting. It features a similar, straddling band alignment as CdSe/ZnS, yet the InP/ZnSe lattice mismatch amounts to a mere 3\%, in stark contrast to the 12\% mismatch between CdSe and ZnS. Although the atomistic origin of blinking in QDs is still debated,\cite{Efros2016} the low-strain, type 1 combination offered by the InP/ZnSe core/shell system could offer a pathway to reconcile non-blinking characteristics with single-photon emission by colloidal QDs.

\begin{figure}
\includegraphics[width=\textwidth]{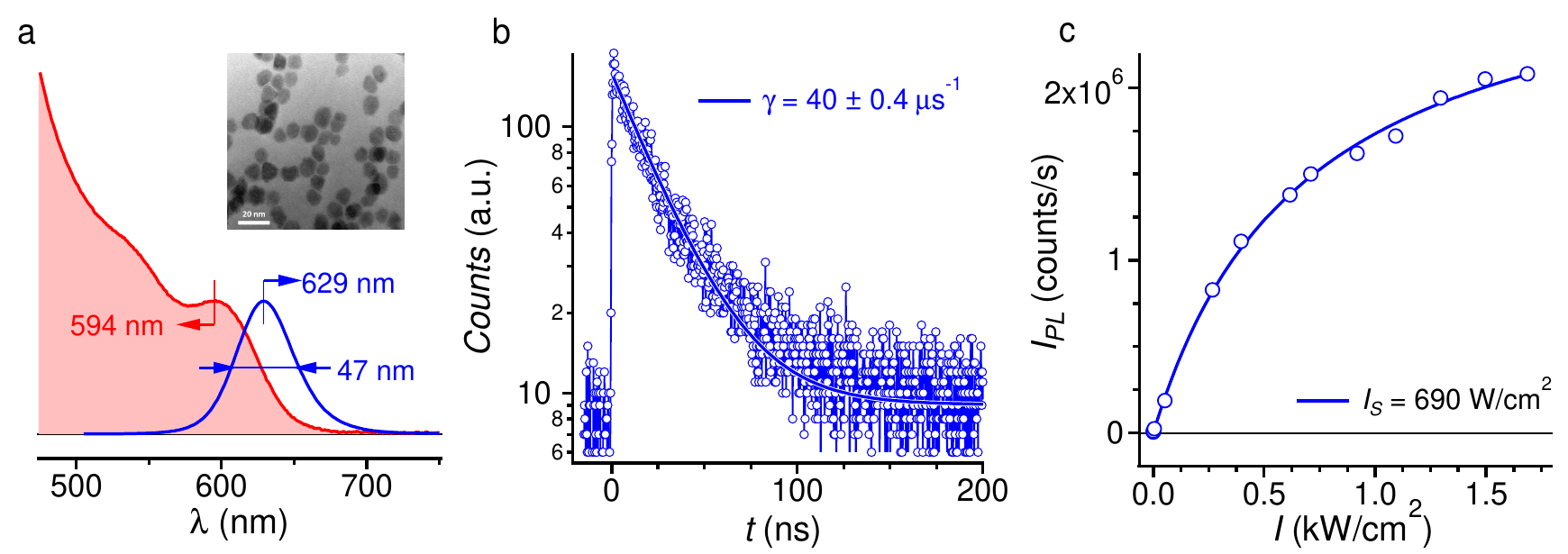}
\caption{(a) Absorption (red) and emission (blue) spectrum of InP/ZnSe quantum dots in solution. Inset: TEM image of the InP/ZnSe quantum dots. (b) Luminescence decay from InP/ZnSe quantum dots in a thin film after pulsed excitation at 445~nm. (c) Luminescence saturation under high intensity cw excitation at 445~nm. \label{Fig1}}
\end{figure}

Here, we study the luminescence of individual InP/ZnSe core/shell QDs (with 3.2-nm cores) synthesized using \text{tris}-diethylaminophosphine as the phosphorous precursor according to a recently published procedure (see Supporting Information S1 for details).\cite{Tessier2016} As shown in Figure \ref{Fig1}a, the resulting InP/ZnSe QDs had an equivalent diameter of about 10~nm (see the inset) and their absorption spectrum exhibited the typical band-edge transition with a maximum at 594~nm. The photoluminescence (PL) in solution featured a concomitant single-peak spectrum around 629~nm, a full-width-at-half maximum (FWHM) of 47 nm, and had a PL quantum yield of 65\%. As shown before, this relatively large FWHM -- Cd-based QDs, $\CPB$ or CdSe nanoplatelets can have an ensemble emission as narrow as 20, $12$, and 10~nm, respectively -- is not a characteristic of the emission of individual InP QDs, yet reflects heterogeneous broadening related to size dispersion.\cite{Ithurria2011,Chen2013,Cui2013, Protesescu2015} 

We further characterized the PL of the InP/ZnSe QDs by analyzing films spincoated on a glass coverslip using the micro-photoluminescence (\uPL) setup described in Supporting Information S2. Focusing on ensemble measurements first, Figure \ref{Fig1}b represents the PL decay obtained after pulsed excitation at 445 nm. The figure includes a best fit of the experimental data to a single exponential function, $\IPL(t)\propto \exp(-\gamma \ t)$, which yields a PL decay rate $\gamma=40\pm0.4~\microsec^{-1}$. Most interestingly, under continuous wave (cw) excitation at 445 nm, the PL intensity $\IPL$ first increases with the excitation intensity $I$, then saturates according to $\IPL\propto I/(I+\IS)$ as expected for an effective 2-level transition (see Figure \ref{Fig1}c). The fitting of the PL intensity yields a saturation intensity $\IS=690\pm40\,\Wcm$. In combination with the photon energy of the cw excitation laser ($\hbar\omega_p$) and the measured decay rate ($\gamma$), this translates into an average QD absorption cross section $\sigma=\hbar\omega_p \gamma/\IS$ of $2.0\pm 0.1~10^{-14}\,\mathrm{cm}^2$. Using the bulk optical constants of InP and ZnSe and the aforementioned dimensions of the core/shell QDs, the absorption cross section of the InP/ZnSe QDs suspended in toluene can be estimated at $2.53~10^{-14}\,\mathrm{cm}^2$ (see Supporting Information S3). The correspondence between both figures indicates that the saturation of the PL reflects the emission saturation of the InP/ZnSe QDs and that they behave as effective 2-level systems.

\begin{figure}[t]
\includegraphics[width=\textwidth]{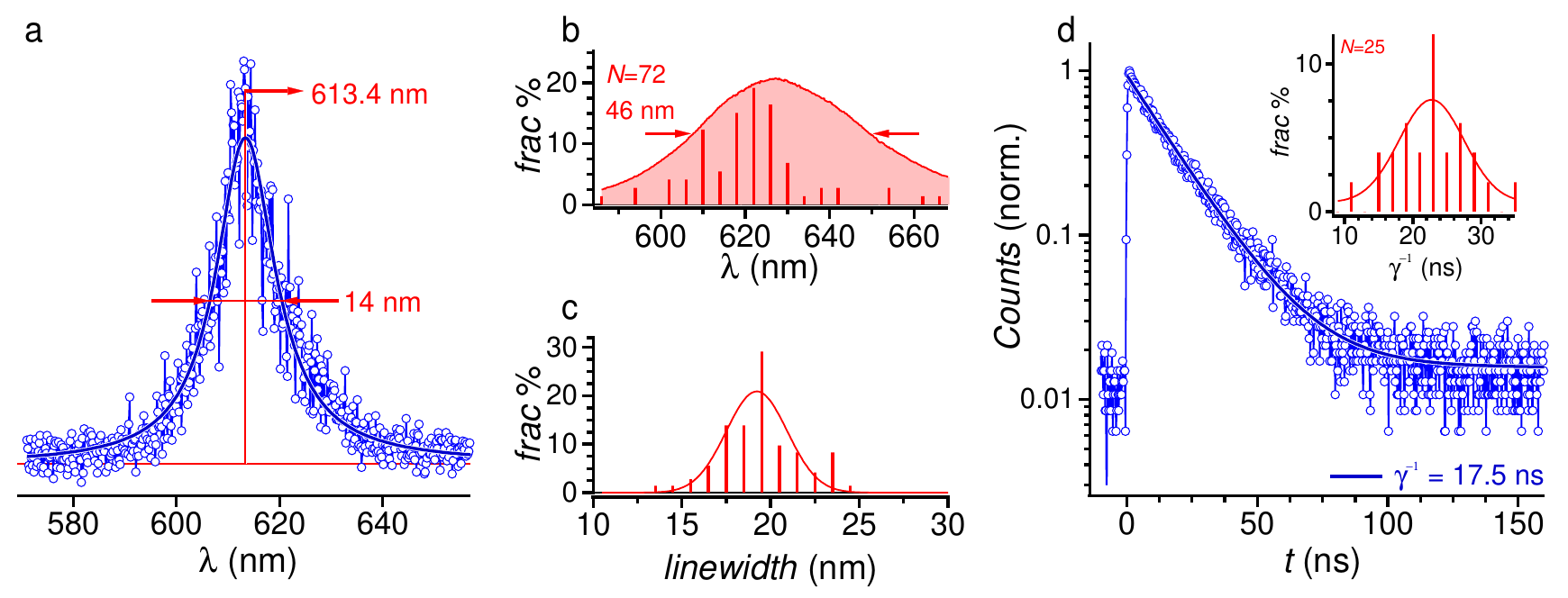}
\caption{(a) Emission spectrum of a single QD fitted to a Lorentzian function. (b) Statistical distribution of the central emission wavelength in an ensemble of 72 single QDs. (c) Statistical distribution of emission linewidth for the same ensemble. (d) Luminescence decay trace of a single QD fitted to a single exponential function with a decay time of 17.5~ns. Inset: statistical distribution of the decay times based on an ensemble of 25 single QDs. \label{Fig2}}
\end{figure}

To address the PL of single InP/ZnSe QDs, a 1-nM solution of QDs in toluene was dropcast on a glass coverslip, producing a QD surface density of $0.25~\text{\textmu m}^{-2}$. Figure~\ref{Fig2}a shows that a single InP/ZnSe QD features a much more narrow emission line than those observed in QD ensembles. A systematic study of 72 single InP/ZnSe QDs showed similarly narrow lines falling within the ensemble emission spectrum (see Figure~\ref{Fig2}b), confirming that the ensemble emission is heterogeneously broadened. The linewidths of single-QD emissions also show some variability (see Figure~\ref{Fig2}c). While the average FWHM amounts to 19.4~nm, the distribution shows a standard deviation of 2.5~nm and the measured values range from 13 to 24~nm. Finally, we found that single InP/ZnSe QDs exhibit a single-exponential PL decay (see Figure \ref{Fig2}d). By investigating an ensemble of 25 QDs, it turned out that the distribution of the decay times $\gamma^{-1}$ can be fitted to a normal law with an average value of $22.5~\ns$ and a standard deviation of $5.4~\ns$ (see inset to Figure \ref{Fig2}d). Similar variations in single QD lifetime have been reported before in the case of CdSe/ZnS QDs.\cite{Fisher2004} We found that the emission lifetime of each QD is strongly correlated to the central energy $E=h\ c/\lambda$ of the emitted photons; no discernible correlation between the spectral linewidth and the photon energy was noticed (see Supporting Information S4).

\begin{figure}[b]
\includegraphics[width=\textwidth]{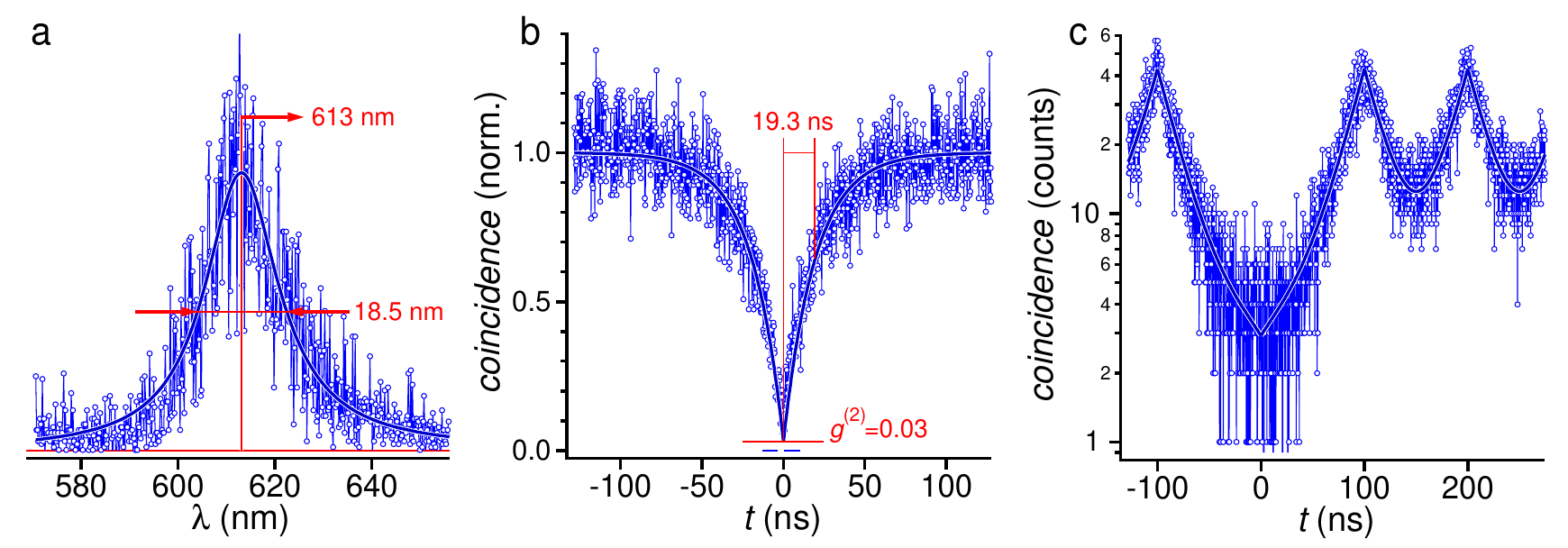}
\caption{(a) Emission spectrum of a single QD. (b) Second-order correlation function $g^{(2)}(t)$ of the QD under cw excitation at 445 nm. (c) Second-order correlation function of the QD under 445-nm (picosecond) pulse excitation at a repetition rate of 10~MHz. The strong antibunching is raw data, without background subtraction nor compensation for intensity differences in start/stop channels. \label{Fig3}}
\end{figure}

Strong photon antibunching was observed from single InP/ZnSe QDs. Figures~\ref{Fig3}a and \ref{Fig3}b show the emission line of a single QD and the corresponding second-order correlation function $g^{(2)}(t)$ measured under cw-laser pumping at 445~nm. The antibunching in Figure \ref{Fig3}b is almost ideal with only a residual background-limited zero-delay value $g^{(2)}(0)=0.03$. We fitted the data to the function $g^{(2)}(t)=1-[1-g^{(2)}(0)] \ \exp(-(\gamma+w_p) \ |t|)$, where $w_p=\sigma I / (\hbar\omega_p)$ is the pumping rate of the QD and $\gamma$ its luminescence decay rate.\cite{Michler2003} The experiment was performed in a weak excitation regime ($w_p\ll\gamma$). The fit yields $(\gamma+w_p)^{-1}\approx \gamma^{-1}=19.3~\ns$ in line with the luminescence decay time statistics measured by direct photon timing (see previous paragraph). We also excited the same QD with 445-nm picosecond pulses at a repetition rate of 10~MHz. No zero-delay peak was seen in the second-order correlation function (see Figure \ref{Fig3}c), even in logarithmic scale, confirming that multi-exciton emission is very efficiently quenched by non-radiative recombination of charge carriers. We measured the $g^{(2)}(0)$ value of 70 single QDs in the cw-pumping regime and found a mean value of 0.19 with a standard deviation of 0.1 (see Supporting Information S5). In each case the antibunching was limited by the residual background noise, not by multi-exciton emission.

\begin{figure}[t]
\includegraphics[width=.5\textwidth]{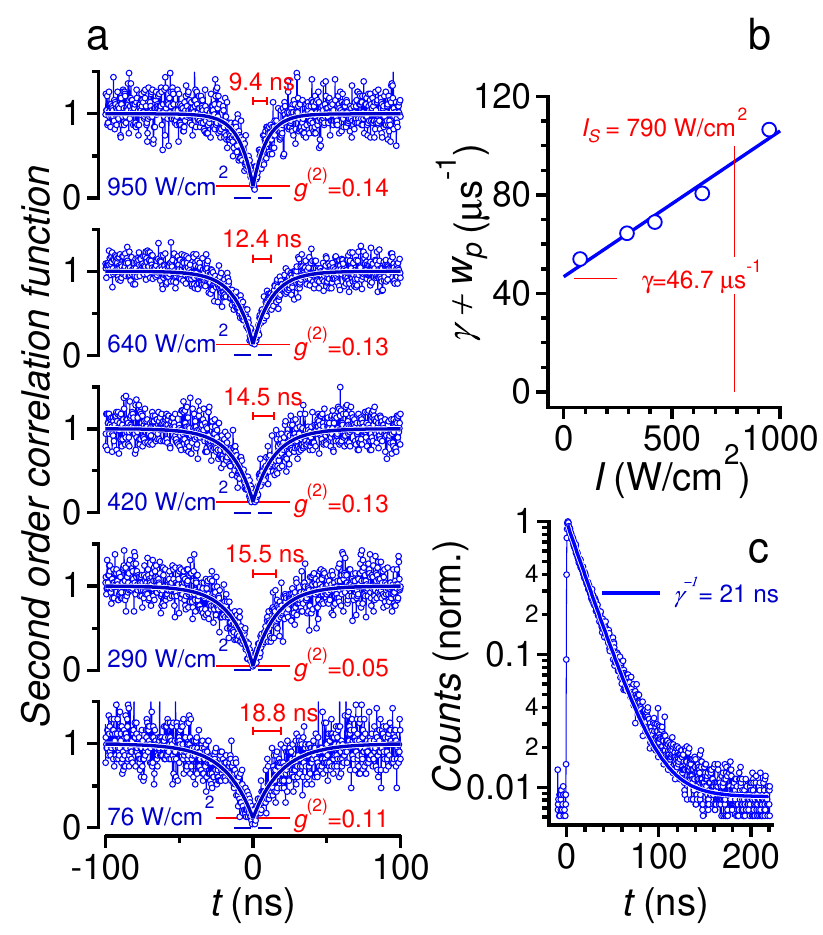}
\caption{(a) Second-order correlation function of a single QD at different cw-pumping intensities at 445~nm. (b) Antibunching width-constant $\gamma+w_p$ as a function of the pump intensity. (c) Luminescence decay of the QD as a function of time.  \label{Fig4}}
\end{figure}

A high photoluminescence quantum yield ($\QY$) and a strong antibunching are key for using InP/ZnSe QDs as integrated single-photon emitters in quantum technology applications. In this perspective, it is crucially important that antibunching be preserved in the saturation regime when $I\gg\IS$ (equivalently, $w_p\gg \gamma$), such that $I/(I+\IS)=w_p/(w_p+\gamma)\approx 1$. If this condition is fulfilled, the emitter can act as a \emph{photon gun}, i.e, a short-pulse excitation with sufficient fluence can completely invert the population of the effective two-level system and trigger the spontaneous emission of a single photon with a probability equal to the $\QY$. In Figure~\ref{Fig4}a, we therefore show second-order correlation traces recorded on a single InP/ZnSe QDs at pump intensities increasing from $76~\Wcm$ to $950~\Wcm$. Over the entire intensity range, the luminescence of the QD preserved a strong antibunching, with a narrowing of the zero-delay dip at higher pump intensity and a stable background-noise limited $g^{(2)}(0)$ between 0.05 and 0.14. By fitting the $g^{(2)}$-functions, we could determine the time constant $(\gamma+w_p)^{-1}$ as indicated in Figure~\ref{Fig4}a. It can be seen that this time constant is always smaller than the emission lifetime -- determined at $\gamma^{-1}=21~\ns$ for this particular QD (see Figure~\ref{Fig4}c), and drops from $18.8~\ns$ to $9.4~\ns$ with increasing pump intensity.

Figure~\ref{Fig4}b represents $\gamma+w_p$ as a function of the pump intensity $I$. In line with the linear dependence of $w_p$ on the pump intensity, $\gamma+w_p$ changes linearly with $I$. A best fit yields a slope $\sigma/(\hbar \omega_p)$ of $5.9~10^4~\mathrm{cm}^2/\mathrm{J}$, from which we obtain the absorption cross-section of the QD as $\sigma=2.0~10^{-14}~\mathrm{cm}^2$. This figure is equal to the ensemble value that was derived from the photoluminescence saturation as plotted in Figure~\ref{Fig1}c. By extrapolating the data in Figure~\ref{Fig4}b to the $I\rightarrow 0$ limit, we find the luminescence decay rate of the QD as $\gamma=47~\microsec^{-1}$, a figure that agrees well with the luminescence decay time $\gamma^{-1}=21~\ns$ obtained in a direct photon-timing experiment (see Figure \ref{Fig4}c). Combining $\sigma$ and $\gamma$, we obtain the saturation intensity of the QD as $\IS=\hbar\omega_p \gamma/\sigma=790~\Wcm$. Hence, the data shown in Figure~\ref{Fig4} demonstrate that strong antibunching is preserved for pumping intensities up to $1.2\times\IS$, with no sign of multi-exciton emission nor significant photo-degradation of the emitter, two essential properties enabling the use of InP/ZnSe QDs as triggered single-photon emitters in quantum optics applications.

In the case of CdSe/ZnS, the persistence of photon antibunching when pumping single QDs above the saturation intensity was attributed to the quenching of radiative recombination of multi-excitons by fast Auger recombination.\cite{Lounis2000} To quantify the rate of Auger recombination, and thus the yield of multi-exciton emission in the case of the InP/ZnSe QDs studied here, we analysed the transient absorbance of an InP/ZnSe dispersion after femtosecond optical pumping. As discussed in the Supporting Information S5, an additional decay component with a time constant of $\sim 70~\ps$ appeared in the transient absorption at high pump intensities. Such a fast component is a typical characteristic of Auger recombination of multi-excitons.\cite{Klimov2000} We therefore interpret the corresponding decay rate $\gamma_{XX}=14.3~\ns^{-1}$ as the combined result of biexciton decay by radiative and non-radiative (Auger) recombination, \textit{i.e.} $\gamma_{XX}=\gamma_{\mathrm{r},XX}+\gamma_{\mathrm{nr},XX}$. Writing the radiative recombination rate of biexcitons\cite{Garcia2011} $\gamma_{\mathrm{r},XX}=4\gamma$ ($\gamma=27.7~\text{\textmu s}^{-1}$ is the decay rate of the fundamental exciton in solution), we can estimate the radiative biexciton quantum yield in InP/ZnSe QDs as: $\QY_{XX}=\gamma_{\mathrm{r},XX}/\gamma_{XX}=0.77\%$. This very low value explains why single-photon emission is retained at high pumping power.

\begin{figure}[t]
\includegraphics[width=\textwidth]{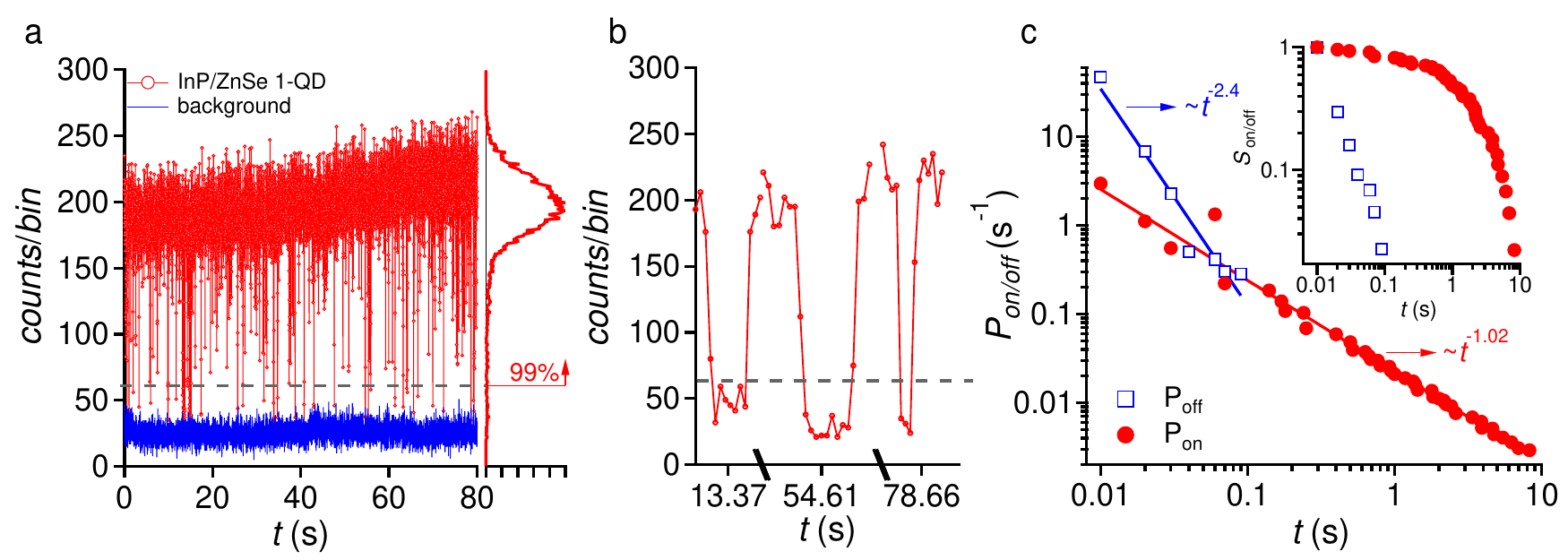}
\caption{(a) Luminescence intensity of the QD in Figure~\ref{Fig4} as a function of time (in counts per 10-ms time bin). The QD is excited with cw 445-nm laser light. (b) Zoom on three unambigously \textit{off} periods. (c) Probability density distributions of the \textit{on} and \textit{off} time periods. The data is fitted to power law distributions $\Ponoff(t)\propto t^{-k_{\text{on (off)}}}$ with $k_{\mathrm{off}}= 2.44$ and $k_{\mathrm{on}}=1.02$. Inset: survival probabilities $\Sonoff(t)=\int_t^\infty \Ponoff(\tau)~\d \tau$ constructed from the data in panel (c). \label{Fig5}}
\end{figure}

Our InP/ZnSe QDs are subject to some modest fluorescence intermittency, a phenomenon commonly known as ``blinking''. At the ensemble level, blinking merely reduces the effective PLQY of the material. In quantum optics applications, however, blinking prohibits the use of single QDs as deterministic single-photon turnstile devices. The data in Figure~\ref{Fig5}a (red line) show the intensity of the light (photon counts per 10-ms time bin) collected from the QD in Figure~\ref{Fig4} as a function of time. The QD was excited with a cw 445-nm laser beam in the low-pumping regime ($I\ll\IS$). The noise background of the measurement is shown in blue. Although some dark periods (no emission, \textit{off} state) can be seen, the histogram of counts displayed on the right of Figure~\ref{Fig5}a indicates that 99\% of time the QD is bright (\textit{on} state) and emits photons at a stable rate. The threshold between \textit{on} and \textit{off} states is indicated by the dashed grey line in Figures~\ref{Fig5}a-b. It was chosen 2-3 standard deviations above the background noise level as can be seen in Figure~\ref{Fig5}b where a zoom on three unambiguously \textit{off} periods is shown. The blinking is very limited and follows a simple \textit{on}/\textit{off} pattern, without any intermediate or grey-state emission, as often reported in the case of CdSe/CdS QDs.\cite{Spinicelli2009} We studied the blinking of 25 different QDs and observed each time the same two-state (\textit{on}/\textit{off}) pattern, with a \textit{on}-state fraction of 95.5\% on average.

To benchmark the fluorescence intermittency in the InP/ZnSe QDs studied here against the extensively studied CdSe/CdS QDs, we define the probability density distributions $\Pon(t)$ and $\Poff(t)$ for the QD to stay in an \textit{on} or an \textit{off} state during the time $t$. The shortest \textit{on} or \textit{off} duration is determined by the selected binning time, which is 10 ms in our measurements, and the longest duration is limited by the emission characteristics of the QD and the time duration of the measurement. The two distributions are calculated from the data in Figure~\ref{Fig5}a in the following way:
\begin{equation}
\Ponoff(t_n) = \frac{\text{Number of ``on'' (``off'') events of duration}~t_n}{\text{Total number of ``on'' (``off'') events}} \times \frac{1}{\Delta_{\mathrm{avg}}(t_n)}
\end{equation}
Here, $\Delta_{\mathrm{avg}}(t_n)=(t_{n+1}-t_{n-1})/2$ is the average of the time intervals to the next shorter and next longer observed events.\cite{Kuno2001} The distributions $\Pon(t)$ (red dots) and $\Poff(t)$ (blue open squares) are plotted in Figure~\ref{Fig5}c in a log-log scale. The linear alignment of the data points in Figure~\ref{Fig5}c indicates that both $\Pon(t)$ and $\Poff(t)$ follow an inverse power law: $\Ponoff(t)\propto t^{-k_{\text{on(off)}}}$ and linear regression (plain lines in Figure~\ref{Fig5}c) shows that $k_{\mathrm{off}}= 2.44$ and $k_{\mathrm{on}}=1.02$. The steeper slope of the \textit{off} events makes \textit{off} periods of 10~ms about 100 times more likely than 100-ms ones. A similar analysis was performed on 25 different QDs. We found an average value of 1.2 for $k_{\mathrm{on}}$ and 2.2 for $k_{\mathrm{off}}$. A value of $k_{\mathrm{off}}>2$ is indicative of efficient blinking suppression and comparable to the values reported in nearly blinking-free CdSe/CdS QD.\cite{Hohng2004,Chen2013} In the inset of Figure~\ref{Fig5}c, we display the survival probabilities $\Sonoff(t)=\int_t^\infty \Ponoff(\tau)~\d \tau$, which represent the probability to find the QD in an \textit{on} or an \textit{off} state for a time longer than $t$. Given the power-law distribution of $\Pon(t)$ and $\Poff(t)$, the survival probabilities $\Sonoff(t)$ scale in principle as $t^{-(k_{\text{on(off)}}-1)}$. However, as $k_{\mathrm{on}}\approx 1$, $S_{\mathrm{on}}(t)$ departs from this power law. The survival probabilities show even more clearly that the \textit{off} periods are remarkably short (10-100~ms) while the \textit{on} periods often last a few seconds or more. As can be seen in the inset of Figure~\ref{Fig5}c, the probability that a QD turns on to remain bright during more than one second is about 50\%. Such long \textit{on} periods are comparable to those reported by Mahler \textit{et al.} in their study on thick-shell non-blinking CdSe/CdS QDs.\cite{Mahler2008}. This observation puts the InP/ZnSe QDs studied here among today's state-of-the-art nearly blinking-free quantum dots. 

In conclusion, we have characterized single InP/ZnSe colloidal QDs showing a unique combination of narrow room-temperature emission spectrum, strong antibunching, mono-exponential luminescence decay, reduced blinking, and photostability. Our single particle spectroscopy study demonstrates that single-photon emission is preserved at saturating intensities, which we attribute to fast non-radiative Auger recombination of multi-excitons. Although these fast multi-exciton recombination rates might limit their use for light amplification, these QDs are ideally suited for quantum optics applications as single-photon turnstile devices. Furthermore, reduced cytotoxicity offers them a prospect in life-science applications. There is scope for further improving the synthesis of the QDs towards even better optical properties. This opens perspectives for future comparative analyses of InP dots with different core/shell sizes, different shell materials, studies at cryogenic temperatures and suitability for nanolithographic deposition.

\begin{acknowledgement}
Z.H. and E.B. acknowledge support by the European Commission via the Marie-Sklodowska Curie action Phonsi (H2020-MSCA-ITN-642656) for funding. Z.H. also acknowledges support from the Belgian Science Policy Office (IAP 7.35, photonics@be), IWT-Vlaanderen (SBO-Lumicor) and Ghent University (GOA n$^{\circ}$ 01G01513).
\end{acknowledgement}

%\begin{suppinfo}
%Materials and chemical synthesis. Micro-photoluminescence setup. Theoretical calculation of absorption cross section. Correlation between emission lifetime, linewidth, and central energy. QD antibunching: $g^{(2)}(0)$ statistics. Pump-probe measurements.
%
%\end{suppinfo}
%\bibliographystyle{achemso}

%\bibliography{manuscript}

\begin{mcitethebibliography}{45}
\providecommand*\natexlab[1]{#1}
\providecommand*\mciteSetBstSublistMode[1]{}
\providecommand*\mciteSetBstMaxWidthForm[2]{}
\providecommand*\mciteBstWouldAddEndPuncttrue
  {\def\EndOfBibitem{\unskip.}}
\providecommand*\mciteBstWouldAddEndPunctfalse
  {\let\EndOfBibitem\relax}
\providecommand*\mciteSetBstMidEndSepPunct[3]{}
\providecommand*\mciteSetBstSublistLabelBeginEnd[3]{}
\providecommand*\EndOfBibitem{}
\mciteSetBstSublistMode{f}
\mciteSetBstMaxWidthForm{subitem}{(\alph{mcitesubitemcount})}
\mciteSetBstSublistLabelBeginEnd
  {\mcitemaxwidthsubitemform\space}
  {\relax}
  {\relax}

\bibitem[Aharonovich et~al.(2016)Aharonovich, Englund, and
  Toth]{Aharonovich2016}
Aharonovich,~I.; Englund,~D.; Toth,~M. \emph{Nat. Photonics} \textbf{2016},
  \emph{10}, 631--641\relax
\mciteBstWouldAddEndPuncttrue
\mciteSetBstMidEndSepPunct{\mcitedefaultmidpunct}
{\mcitedefaultendpunct}{\mcitedefaultseppunct}\relax
\EndOfBibitem
\bibitem[Beveratos et~al.(2002)Beveratos, Brouri, Gacoin, Villing, Poizat, and
  Grangier]{Beveratos2002}
Beveratos,~A.; Brouri,~R.; Gacoin,~T.; Villing,~A.; Poizat,~J.-P.; Grangier,~P.
  \emph{Phys. Rev. Lett.} \textbf{2002}, \emph{89}, 187901\relax
\mciteBstWouldAddEndPuncttrue
\mciteSetBstMidEndSepPunct{\mcitedefaultmidpunct}
{\mcitedefaultendpunct}{\mcitedefaultseppunct}\relax
\EndOfBibitem
\bibitem[Heindel et~al.(2012)Heindel, Kessler, Rau, Schneider, F\"{u}rst,
  Hargart, Schulz, Eichfelder, Ro{\ss}bach, Nauerth, Lermer, Weier, Jetter,
  Kamp, Reitzenstein, H\"{o}fling, Michler, Weinfurter, and
  Forchel]{Heindel2012}
Heindel,~T. et~al.  \emph{New J. Phys.} \textbf{2012}, \emph{14}, 083001\relax
\mciteBstWouldAddEndPuncttrue
\mciteSetBstMidEndSepPunct{\mcitedefaultmidpunct}
{\mcitedefaultendpunct}{\mcitedefaultseppunct}\relax
\EndOfBibitem
\bibitem[Broome et~al.(2010)Broome, Fedrizzi, Lanyon, Kassal, Aspuru-Guzik, and
  White]{Broome2010}
Broome,~M.~A.; Fedrizzi,~A.; Lanyon,~B.~P.; Kassal,~I.; Aspuru-Guzik,~A.;
  White,~A.~G. \emph{Phys. Rev. Lett.} \textbf{2010}, \emph{104}, 153602\relax
\mciteBstWouldAddEndPuncttrue
\mciteSetBstMidEndSepPunct{\mcitedefaultmidpunct}
{\mcitedefaultendpunct}{\mcitedefaultseppunct}\relax
\EndOfBibitem
\bibitem[Tillmann et~al.(2013)Tillmann, Daki\'{c}, Heilmann, Nolte, Szameit,
  and Walther]{Tillmann2013}
Tillmann,~M.; Daki\'{c},~B.; Heilmann,~R.; Nolte,~S.; Szameit,~A.; Walther,~P.
  \emph{Nat. Photonics} \textbf{2013}, \emph{7}, 540--544\relax
\mciteBstWouldAddEndPuncttrue
\mciteSetBstMidEndSepPunct{\mcitedefaultmidpunct}
{\mcitedefaultendpunct}{\mcitedefaultseppunct}\relax
\EndOfBibitem
\bibitem[Motes et~al.(2015)Motes, Olson, Rabeaux, Dowling, Olson, and
  Rohde]{Motes2015}
Motes,~K.~R.; Olson,~J.~P.; Rabeaux,~E.~J.; Dowling,~J.~P.; Olson,~S.~J.;
  Rohde,~P.~P. \emph{Phys. Rev. Lett.} \textbf{2015}, \emph{114}, 170802\relax
\mciteBstWouldAddEndPuncttrue
\mciteSetBstMidEndSepPunct{\mcitedefaultmidpunct}
{\mcitedefaultendpunct}{\mcitedefaultseppunct}\relax
\EndOfBibitem
\bibitem[Aharonovich and Neu(2014)Aharonovich, and Neu]{Aharonovich2014}
Aharonovich,~I.; Neu,~E. \emph{Adv. Opt. Mater.} \textbf{2014}, \emph{2},
  911--928\relax
\mciteBstWouldAddEndPuncttrue
\mciteSetBstMidEndSepPunct{\mcitedefaultmidpunct}
{\mcitedefaultendpunct}{\mcitedefaultseppunct}\relax
\EndOfBibitem
\bibitem[Morfa et~al.(2012)Morfa, Gibson, Karg, Karle, Greentree, Mulvaney, and
  Tomljenovic-Hanic]{Morfa2012}
Morfa,~A.~J.; Gibson,~B.~C.; Karg,~M.; Karle,~T.~J.; Greentree,~A.~D.;
  Mulvaney,~P.; Tomljenovic-Hanic,~S. \emph{Nano Lett.} \textbf{2012},
  \emph{12}, 949--954\relax
\mciteBstWouldAddEndPuncttrue
\mciteSetBstMidEndSepPunct{\mcitedefaultmidpunct}
{\mcitedefaultendpunct}{\mcitedefaultseppunct}\relax
\EndOfBibitem
\bibitem[Utikal et~al.(2014)Utikal, Eichhammer, Petersen, Renn, G\"{o}tzinger,
  and Sandoghdar]{Utikal2014}
Utikal,~T.; Eichhammer,~E.; Petersen,~L.; Renn,~A.; G\"{o}tzinger,~S.;
  Sandoghdar,~V. \emph{Nat. Commun.} \textbf{2014}, \emph{5}, 3627\relax
\mciteBstWouldAddEndPuncttrue
\mciteSetBstMidEndSepPunct{\mcitedefaultmidpunct}
{\mcitedefaultendpunct}{\mcitedefaultseppunct}\relax
\EndOfBibitem
\bibitem[Ma et~al.(2015)Ma, Hartmann, Baldwin, Doorn, and Htoon]{Ma2015}
Ma,~X.; Hartmann,~N.~F.; Baldwin,~J.~K.; Doorn,~S.~K.; Htoon,~H. \emph{Nature
  nanotechnology} \textbf{2015}, \emph{10}, 671--675\relax
\mciteBstWouldAddEndPuncttrue
\mciteSetBstMidEndSepPunct{\mcitedefaultmidpunct}
{\mcitedefaultendpunct}{\mcitedefaultseppunct}\relax
\EndOfBibitem
\bibitem[He et~al.(2015)He, Clark, Schaibley, He, Chen, Wei, Ding, Zhang, Yao,
  Xu, Lu, and Pan]{He2015}
He,~Y.-M.; Clark,~G.; Schaibley,~J.~R.; He,~Y.; Chen,~M.-C.; Wei,~Y.-J.;
  Ding,~X.; Zhang,~Q.; Yao,~W.; Xu,~X.; Lu,~C.-Y.; Pan,~J.-W. \emph{Nat.
  Nanotechnol.} \textbf{2015}, \emph{10}, 497--502\relax
\mciteBstWouldAddEndPuncttrue
\mciteSetBstMidEndSepPunct{\mcitedefaultmidpunct}
{\mcitedefaultendpunct}{\mcitedefaultseppunct}\relax
\EndOfBibitem
\bibitem[Buckley et~al.(2012)Buckley, Rivoire, and
  Vu\v{c}kovi\'{c}]{Buckley2012}
Buckley,~S.; Rivoire,~K.; Vu\v{c}kovi\'{c},~J. \emph{Rep. Prog. Phys.}
  \textbf{2012}, \emph{75}, 126503\relax
\mciteBstWouldAddEndPuncttrue
\mciteSetBstMidEndSepPunct{\mcitedefaultmidpunct}
{\mcitedefaultendpunct}{\mcitedefaultseppunct}\relax
\EndOfBibitem
\bibitem[Patel et~al.(2010)Patel, Bennett, Farrer, Nicoll, Ritchie, and
  Shields]{Patel2010}
Patel,~R.~B.; Bennett,~A.~J.; Farrer,~I.; Nicoll,~C.~A.; Ritchie,~D.~A.;
  Shields,~A.~J. \emph{Nat. Photonics} \textbf{2010}, \emph{4}, 632--635\relax
\mciteBstWouldAddEndPuncttrue
\mciteSetBstMidEndSepPunct{\mcitedefaultmidpunct}
{\mcitedefaultendpunct}{\mcitedefaultseppunct}\relax
\EndOfBibitem
\bibitem[Sapienza et~al.(2015)Sapienza, Davan\c{c}o, Badolato, and
  Srinivasan]{Sapienza2015}
Sapienza,~L.; Davan\c{c}o,~M.; Badolato,~A.; Srinivasan,~K. \emph{Nat. Commun.}
  \textbf{2015}, \emph{6}, 7833\relax
\mciteBstWouldAddEndPuncttrue
\mciteSetBstMidEndSepPunct{\mcitedefaultmidpunct}
{\mcitedefaultendpunct}{\mcitedefaultseppunct}\relax
\EndOfBibitem
\bibitem[Brokmann et~al.(2004)Brokmann, Giacobino, Dahan, and
  Hermier]{Brokmann2004}
Brokmann,~X.; Giacobino,~E.; Dahan,~M.; Hermier,~J.-P. \emph{Appl. Phys. Lett.}
  \textbf{2004}, \emph{85}, 712--714\relax
\mciteBstWouldAddEndPuncttrue
\mciteSetBstMidEndSepPunct{\mcitedefaultmidpunct}
{\mcitedefaultendpunct}{\mcitedefaultseppunct}\relax
\EndOfBibitem
\bibitem[Brokmann et~al.(2004)Brokmann, Messin, Desbiolles, Giacobino, Dahan,
  and Hermier]{Brokmann2004a}
Brokmann,~X.; Messin,~G.; Desbiolles,~P.; Giacobino,~E.; Dahan,~M.;
  Hermier,~J.-P. \emph{New J. Phys.} \textbf{2004}, \emph{6}, 99\relax
\mciteBstWouldAddEndPuncttrue
\mciteSetBstMidEndSepPunct{\mcitedefaultmidpunct}
{\mcitedefaultendpunct}{\mcitedefaultseppunct}\relax
\EndOfBibitem
\bibitem[Pisanello et~al.(2010)Pisanello, Martiradonna, Lem\'{e}nager,
  Spinicelli, Fiore, Manna, Hermier, Cingolani, Giacobino, De~Vittorio, and
  Bramati]{Pisanello2010}
Pisanello,~F.; Martiradonna,~L.; Lem\'{e}nager,~G.; Spinicelli,~P.; Fiore,~A.;
  Manna,~L.; Hermier,~J.-P.; Cingolani,~R.; Giacobino,~E.; De~Vittorio,~M.;
  Bramati,~A. \emph{Appl. Phys. Lett.} \textbf{2010}, \emph{96}, 033101\relax
\mciteBstWouldAddEndPuncttrue
\mciteSetBstMidEndSepPunct{\mcitedefaultmidpunct}
{\mcitedefaultendpunct}{\mcitedefaultseppunct}\relax
\EndOfBibitem
\bibitem[Pisanello et~al.(2013)Pisanello, Lem\'{e}nager, Martiradonna, Carbone,
  Vezzoli, Desfonds, Cozzoli, Hermier, Giacobino, Cingolani, De~Vittorio, and
  Bramati]{Pisanello2013}
Pisanello,~F.; Lem\'{e}nager,~G.; Martiradonna,~L.; Carbone,~L.; Vezzoli,~S.;
  Desfonds,~P.; Cozzoli,~P.~D.; Hermier,~J.-P.; Giacobino,~E.; Cingolani,~R.;
  De~Vittorio,~M.; Bramati,~A. \emph{Adv. Mater.} \textbf{2013}, \emph{25},
  1974--1980\relax
\mciteBstWouldAddEndPuncttrue
\mciteSetBstMidEndSepPunct{\mcitedefaultmidpunct}
{\mcitedefaultendpunct}{\mcitedefaultseppunct}\relax
\EndOfBibitem
\bibitem[Michler et~al.(2000)Michler, Imamoglu, Mason, Carson, Strouse, and
  Buratto]{Michler2000}
Michler,~P.; Imamoglu,~A.; Mason,~M.~D.; Carson,~P.~J.; Strouse,~G.~F.;
  Buratto,~S.~K. \emph{Nature} \textbf{2000}, \emph{406}, 968--970\relax
\mciteBstWouldAddEndPuncttrue
\mciteSetBstMidEndSepPunct{\mcitedefaultmidpunct}
{\mcitedefaultendpunct}{\mcitedefaultseppunct}\relax
\EndOfBibitem
\bibitem[Lounis et~al.(2000)Lounis, Bechtel, Gerion, Alivisatos, and
  Moerner]{Lounis2000}
Lounis,~B.; Bechtel,~H.~A.; Gerion,~D.; Alivisatos,~P.; Moerner,~W.~E.
  \emph{Chem. Phys. Lett.} \textbf{2000}, \emph{329}, 399--404\relax
\mciteBstWouldAddEndPuncttrue
\mciteSetBstMidEndSepPunct{\mcitedefaultmidpunct}
{\mcitedefaultendpunct}{\mcitedefaultseppunct}\relax
\EndOfBibitem
\bibitem[Nirmal et~al.(1996)Nirmal, Dabbousi, Bawendi, Macklin, Trautman,
  Harris, and Brus]{Nirmal1996}
Nirmal,~M.; Dabbousi,~B.~O.; Bawendi,~M.~G.; Macklin,~J.~J.; Trautman,~J.~K.;
  Harris,~T.~D.; Brus,~L.~E. \emph{Nature} \textbf{1996}, \emph{383},
  802--804\relax
\mciteBstWouldAddEndPuncttrue
\mciteSetBstMidEndSepPunct{\mcitedefaultmidpunct}
{\mcitedefaultendpunct}{\mcitedefaultseppunct}\relax
\EndOfBibitem
\bibitem[Chen et~al.(2008)Chen, Vela, Htoon, Casson, Werder, Bussian, Klimov,
  and Hollingsworth]{Chen2008}
Chen,~Y.; Vela,~J.; Htoon,~H.; Casson,~J.~L.; Werder,~D.~J.; Bussian,~D.~A.;
  Klimov,~V.~I.; Hollingsworth,~J.~A. \emph{J. Am. Chem. Soc.} \textbf{2008},
  \emph{130}, 5026--5027\relax
\mciteBstWouldAddEndPuncttrue
\mciteSetBstMidEndSepPunct{\mcitedefaultmidpunct}
{\mcitedefaultendpunct}{\mcitedefaultseppunct}\relax
\EndOfBibitem
\bibitem[Mahler et~al.(2008)Mahler, Spinicelli, Buil, Quelin, Hermier, and
  Dubertret]{Mahler2008}
Mahler,~B.; Spinicelli,~P.; Buil,~S.; Quelin,~X.; Hermier,~J.-P.; Dubertret,~B.
  \emph{Nat. Mater.} \textbf{2008}, \emph{7}, 659--664\relax
\mciteBstWouldAddEndPuncttrue
\mciteSetBstMidEndSepPunct{\mcitedefaultmidpunct}
{\mcitedefaultendpunct}{\mcitedefaultseppunct}\relax
\EndOfBibitem
\bibitem[Park et~al.(2014)Park, Bae, Padilha, Pietryga, and Klimov]{Park2014}
Park,~Y.-S.; Bae,~W.~K.; Padilha,~L.~A.; Pietryga,~J.~M.; Klimov,~V.~I.
  \emph{Nano Lett.} \textbf{2014}, \emph{14}, 396--402\relax
\mciteBstWouldAddEndPuncttrue
\mciteSetBstMidEndSepPunct{\mcitedefaultmidpunct}
{\mcitedefaultendpunct}{\mcitedefaultseppunct}\relax
\EndOfBibitem
\bibitem[Yuan et~al.(2009)Yuan, Yu, Ko, Huang, and Tang]{Yuan2009}
Yuan,~C.~T.; Yu,~P.; Ko,~H.~C.; Huang,~J.; Tang,~J. \emph{ACS Nano}
  \textbf{2009}, \emph{3}, 3051--3056\relax
\mciteBstWouldAddEndPuncttrue
\mciteSetBstMidEndSepPunct{\mcitedefaultmidpunct}
{\mcitedefaultendpunct}{\mcitedefaultseppunct}\relax
\EndOfBibitem
\bibitem[Naiki et~al.(2011)Naiki, Masuo, Machida, and Itaya]{Naiki2011}
Naiki,~H.; Masuo,~S.; Machida,~S.; Itaya,~A. \emph{J. Phys. Chem. C}
  \textbf{2011}, \emph{115}, 23299--23304\relax
\mciteBstWouldAddEndPuncttrue
\mciteSetBstMidEndSepPunct{\mcitedefaultmidpunct}
{\mcitedefaultendpunct}{\mcitedefaultseppunct}\relax
\EndOfBibitem
\bibitem[Protesescu et~al.(2015)Protesescu, Yakunin, Bodnarchuk, Krieg, Caputo,
  Hendon, Yang, Walsh, and Kovalenko]{Protesescu2015}
Protesescu,~L.; Yakunin,~S.; Bodnarchuk,~M.~I.; Krieg,~F.; Caputo,~R.;
  Hendon,~C.~H.; Yang,~R.~X.; Walsh,~A.; Kovalenko,~M.~V. \emph{Nano Lett.}
  \textbf{2015}, \emph{15}, 3692--3696\relax
\mciteBstWouldAddEndPuncttrue
\mciteSetBstMidEndSepPunct{\mcitedefaultmidpunct}
{\mcitedefaultendpunct}{\mcitedefaultseppunct}\relax
\EndOfBibitem
\bibitem[Park et~al.(2015)Park, Guo, Makarov, and Klimov]{Park2015}
Park,~Y.-S.; Guo,~S.; Makarov,~N.~S.; Klimov,~V.~I. \emph{ACS Nano}
  \textbf{2015}, \emph{9}, 10386--10393\relax
\mciteBstWouldAddEndPuncttrue
\mciteSetBstMidEndSepPunct{\mcitedefaultmidpunct}
{\mcitedefaultendpunct}{\mcitedefaultseppunct}\relax
\EndOfBibitem
\bibitem[Raino et~al.(2016)Raino, Nedelcu, Protesescu, Bodnarchuk, Kovalenko,
  Mahrt, and Stoeferle]{Raino2016}
Raino,~G.; Nedelcu,~G.; Protesescu,~L.; Bodnarchuk,~M.~I.; Kovalenko,~M.~V.;
  Mahrt,~R.~F.; Stoeferle,~T. \emph{ACS Nano} \textbf{2016}, \emph{10},
  2485--2490\relax
\mciteBstWouldAddEndPuncttrue
\mciteSetBstMidEndSepPunct{\mcitedefaultmidpunct}
{\mcitedefaultendpunct}{\mcitedefaultseppunct}\relax
\EndOfBibitem
\bibitem[Micic et~al.(1994)Micic, Curtis, Jones, Sprague, and Nozik]{micic1994}
Micic,~O.~I.; Curtis,~C.~J.; Jones,~K.~M.; Sprague,~J.~R.; Nozik,~A.~J.
  \emph{J. Phys. Chem.} \textbf{1994}, \emph{98}, 4966--4969\relax
\mciteBstWouldAddEndPuncttrue
\mciteSetBstMidEndSepPunct{\mcitedefaultmidpunct}
{\mcitedefaultendpunct}{\mcitedefaultseppunct}\relax
\EndOfBibitem
\bibitem[Song et~al.(2013)Song, Lee, Lee, Jang, Choi, Choi, and Yang]{Song2013}
Song,~W.-S.; Lee,~H.-S.; Lee,~J.~C.; Jang,~D.~S.; Choi,~Y.; Choi,~M.; Yang,~H.
  \emph{J. Nanopart. Res.} \textbf{2013}, \emph{15}, 1750\relax
\mciteBstWouldAddEndPuncttrue
\mciteSetBstMidEndSepPunct{\mcitedefaultmidpunct}
{\mcitedefaultendpunct}{\mcitedefaultseppunct}\relax
\EndOfBibitem
\bibitem[Tessier et~al.(2015)Tessier, Dupont, De~Nolf, De~Roo, and
  Hens]{Tessier2015}
Tessier,~M.~D.; Dupont,~D.; De~Nolf,~K.; De~Roo,~J.; Hens,~Z. \emph{Chem.
  Mater.} \textbf{2015}, \emph{27}, 4893--4898\relax
\mciteBstWouldAddEndPuncttrue
\mciteSetBstMidEndSepPunct{\mcitedefaultmidpunct}
{\mcitedefaultendpunct}{\mcitedefaultseppunct}\relax
\EndOfBibitem
\bibitem[Efros and Nesbitt(2016)Efros, and Nesbitt]{Efros2016}
Efros,~A.~L.; Nesbitt,~D.~J. \emph{Nat. Nanotechnol.} \textbf{2016}, \emph{11},
  661--671\relax
\mciteBstWouldAddEndPuncttrue
\mciteSetBstMidEndSepPunct{\mcitedefaultmidpunct}
{\mcitedefaultendpunct}{\mcitedefaultseppunct}\relax
\EndOfBibitem
\bibitem[Tessier et~al.(2016)Tessier, De~Nolf, Dupont, Sinnaeve, De~Roo, and
  Hens]{Tessier2016}
Tessier,~M.~D.; De~Nolf,~K.; Dupont,~D.; Sinnaeve,~D.; De~Roo,~J.; Hens,~Z.
  \emph{J. Am. Chem. Soc.} \textbf{2016}, \emph{138}, 5923--5929\relax
\mciteBstWouldAddEndPuncttrue
\mciteSetBstMidEndSepPunct{\mcitedefaultmidpunct}
{\mcitedefaultendpunct}{\mcitedefaultseppunct}\relax
\EndOfBibitem
\bibitem[Ithurria et~al.(2011)Ithurria, Tessier, Mahler, Lobo, Dubertret, and
  Efros]{Ithurria2011}
Ithurria,~S.; Tessier,~M.~D.; Mahler,~B.; Lobo,~R. P. S.~M.; Dubertret,~B.;
  Efros,~A.~L. \emph{Nat. Mater.} \textbf{2011}, \emph{10}, 936--941\relax
\mciteBstWouldAddEndPuncttrue
\mciteSetBstMidEndSepPunct{\mcitedefaultmidpunct}
{\mcitedefaultendpunct}{\mcitedefaultseppunct}\relax
\EndOfBibitem
\bibitem[Chen et~al.(2013)Chen, Zhao, Chauhan, Cui, Wong, Harris, Wei, Han,
  Fukumura, Jain, and Bawendi]{Chen2013}
Chen,~O.; Zhao,~J.; Chauhan,~V.~P.; Cui,~J.; Wong,~C.; Harris,~D.~K.; Wei,~H.;
  Han,~H.-S.; Fukumura,~D.; Jain,~R.~K.; Bawendi,~M.~G. \emph{Nat. Mater.}
  \textbf{2013}, \emph{12}, 445--451\relax
\mciteBstWouldAddEndPuncttrue
\mciteSetBstMidEndSepPunct{\mcitedefaultmidpunct}
{\mcitedefaultendpunct}{\mcitedefaultseppunct}\relax
\EndOfBibitem
\bibitem[Cui et~al.(2013)Cui, Beyler, Marshall, Chen, Harris, Wanger, Brokmann,
  and Bawendi]{Cui2013}
Cui,~J.; Beyler,~A.~P.; Marshall,~L.~F.; Chen,~O.; Harris,~D.~K.;
  Wanger,~D.~D.; Brokmann,~X.; Bawendi,~M.~G. \emph{Nat. Chem.} \textbf{2013},
  \emph{5}, 602--606\relax
\mciteBstWouldAddEndPuncttrue
\mciteSetBstMidEndSepPunct{\mcitedefaultmidpunct}
{\mcitedefaultendpunct}{\mcitedefaultseppunct}\relax
\EndOfBibitem
\bibitem[Fisher et~al.(2004)Fisher, Eisler, Stott, and Bawendi]{Fisher2004}
Fisher,~B.~R.; Eisler,~H.~J.; Stott,~N.~E.; Bawendi,~M.~G. \emph{J. Phys. Chem.
  B} \textbf{2004}, \emph{108}, 143--148\relax
\mciteBstWouldAddEndPuncttrue
\mciteSetBstMidEndSepPunct{\mcitedefaultmidpunct}
{\mcitedefaultendpunct}{\mcitedefaultseppunct}\relax
\EndOfBibitem
\bibitem[Michler(2003)]{Michler2003}
Michler,~P. In \emph{Single Quantum Dots: Fundamentals, Applications, and New
  Concepts}; Michler,~P., Ed.; Springer Berlin Heidelberg: Berlin, Heidelberg,
  2003; Chapter 8, pp 315--347\relax
\mciteBstWouldAddEndPuncttrue
\mciteSetBstMidEndSepPunct{\mcitedefaultmidpunct}
{\mcitedefaultendpunct}{\mcitedefaultseppunct}\relax
\EndOfBibitem
\bibitem[Klimov et~al.(2000)Klimov, Mikhailovsky, McBranch, Leatherdale, and
  Bawendi]{Klimov2000}
Klimov,~V.~I.; Mikhailovsky,~A.~A.; McBranch,~D.~W.; Leatherdale,~C.~A.;
  Bawendi,~M.~G. \emph{Science} \textbf{2000}, \emph{287}, 1011--1013\relax
\mciteBstWouldAddEndPuncttrue
\mciteSetBstMidEndSepPunct{\mcitedefaultmidpunct}
{\mcitedefaultendpunct}{\mcitedefaultseppunct}\relax
\EndOfBibitem
\bibitem[Garcia-Santamaria et~al.(2011)Garcia-Santamaria, Brovelli, Viswanatha,
  Hollingsworth, Htoon, Crooker, and Klimov]{Garcia2011}
Garcia-Santamaria,~F.; Brovelli,~S.; Viswanatha,~R.; Hollingsworth,~J.~A.;
  Htoon,~H.; Crooker,~S.~A.; Klimov,~V.~I. \emph{Nano Lett.} \textbf{2011},
  \emph{11}, 687--693\relax
\mciteBstWouldAddEndPuncttrue
\mciteSetBstMidEndSepPunct{\mcitedefaultmidpunct}
{\mcitedefaultendpunct}{\mcitedefaultseppunct}\relax
\EndOfBibitem
\bibitem[Spinicelli et~al.(2009)Spinicelli, Buil, Qu\'{e}lin, Mahler,
  Dubertret, and Hermier]{Spinicelli2009}
Spinicelli,~P.; Buil,~S.; Qu\'{e}lin,~X.; Mahler,~B.; Dubertret,~B.;
  Hermier,~J.-P. \emph{Phys. Rev. Lett.} \textbf{2009}, \emph{102},
  136801\relax
\mciteBstWouldAddEndPuncttrue
\mciteSetBstMidEndSepPunct{\mcitedefaultmidpunct}
{\mcitedefaultendpunct}{\mcitedefaultseppunct}\relax
\EndOfBibitem
\bibitem[Kuno et~al.(2001)Kuno, Fromm, Hamann, Gallagher, and
  Nesbitt]{Kuno2001}
Kuno,~M.; Fromm,~D.~P.; Hamann,~H.~F.; Gallagher,~A.; Nesbitt,~D.~J. \emph{J.
  Chem. Phys.} \textbf{2001}, \emph{115}, 1028--1040\relax
\mciteBstWouldAddEndPuncttrue
\mciteSetBstMidEndSepPunct{\mcitedefaultmidpunct}
{\mcitedefaultendpunct}{\mcitedefaultseppunct}\relax
\EndOfBibitem
\bibitem[Hohng and Ha(2004)Hohng, and Ha]{Hohng2004}
Hohng,~S.; Ha,~T. \emph{J. Am. Chem. Soc.} \textbf{2004}, \emph{126},
  1324--1325\relax
\mciteBstWouldAddEndPuncttrue
\mciteSetBstMidEndSepPunct{\mcitedefaultmidpunct}
{\mcitedefaultendpunct}{\mcitedefaultseppunct}\relax
\EndOfBibitem
\end{mcitethebibliography}

\providecommand*\mcitethebibliography{\thebibliography}
\csname @ifundefined\endcsname{endmcitethebibliography}
  {\let\endmcitethebibliography\endthebibliography}{}

\end{document}